\documentclass[twocolumn]{aastex62}
\usepackage[utf8]{inputenc}
\usepackage{natbib, tabularx,hyperref}
\usepackage[caption=false]{subfig}
\usepackage{amsmath, xspace}
\date{March 11, 2021}

\shorttitle{Predicting Primordial Stars}
\shortauthors{Wells and Norman}

\newcommand{\msun}{M$_\odot$\xspace}
\newcommand{\pii}{Pop II\xspace}
\newcommand{\piii}{Pop III\xspace}
\newcommand{\enzo}{{\tt Enzo}\xspace}
\newcommand{\sss}{{\tt StarFind}\xspace}

\begin{document}

\title{Predicting Localized Primordial Star Formation With Deep Convolutional Neural Networks}

\author{Azton I. Wells}
\email{aiwells@ucsd.edu}
\affil{Center for Astrophysics and Space Sciences\\
University of California, San Diego, La Jolla, CA, 92093}
\author{Michael L. Norman}
\affil{Center for Astrophysics and Space Sciences\\
University of California, San Diego, La Jolla, CA, 92093}
\affil{San Diego Supercomputer Center\\
University of California, San Diego, La Jolla, CA, 92093}

\begin{abstract}
We investigate applying 3D deep convolutional neural networks as fast surrogate models of the formation and feedback effects of primordial stars in hydrodynamic cosmological simulations of the first galaxies. Here, we present the surrogate model to predict localized primordial star formation; the feedback model will be presented in a subsequent paper. The star formation prediction model consists of two sub-models: the first is a 3D volume classifier that predicts which (10 comoving kpc)$^3$  volumes will host star formation, followed by a 3D Inception-based U-net voxel segmentation model that predicts which voxels will form primordial stars.  We find that the combined model predicts primordial star forming volumes with high skill, with $F_1 >0.995$ and true skill score $>0.994$. The star formation is localized within the volume to $\lesssim5^3$~voxels ($\sim1.6$~comoving kpc$^3$) with $F_1>0.399$ and true skill score $>0.857$.  Applied to simulations with low spatial resolution, the model predicts star forming regions in the same locations and at similar redshifts as sites in resolved full-physics simulations that explicitly model primordial star formation and feedback. When applied to simulations with lower mass resolution, we find that the model predicts star forming regions at later redshift due to delayed structure formation resulting from lower mass resolution.  Our model predicts primordial star formation without halo finding, so will be useful in spatially under-resolved simulations that cannot resolve primordial star forming halos. To our knowledge, this is the first model that can predict primordial star forming regions that match highly-resolved cosmological simulations.  
\end{abstract}

\section{Introduction}
Despite the rise of petascale computing, astrophysical simulations continue to push the limits of the most advanced supercomputers.  While dark matter (DM) or hydrodynamic gas-only simulations can now simulate massive volumes~(e.g., \citealt{vogels2014}), the inclusion of more complete physical models, such as resolved star formation and feedback (SFF) processes, severely limits the volume of feasible simulations~\citep{Hopkins2018, smith2015}.  
When attempting precision modelling of high redshift galaxy formation with associated SFF, one must usually choose whether to simulate the primordial star (\piii) formation era, or to adopt a simplification scheme to compensate for the lack thereof. Here, we investigate a third option: use deep learning to train a surrogate model with data from simulations which resolve the primordial star formation and feedback processes directly. Then, use the surrogate model in inference to predict the coarse-grained effects of primordial stellar feedback in a large volume cosmological simulation. With SFF modeled rather than simulated, spatial resolution and timestepping requirements are greatly relaxed, thus accelerating the time to solution without sacrificing essential feedback effects such as chemical enrichment by primordial supernovae. In the following, we motivate this approach by reviewing primordial SFF as well as recent applications of deep learning to astrophysical applications.  

Starting at $z\simeq30$, \piii stars begin to form from pristine (H, He, H$_2$) gas in mini-halos with virial mass $M_{vir}\gtrsim10^{5-6.5}$\msun\citep{bromm_review}.  After formation, \piii stars may directly collapse to black holes (BHs), or if formed in the right mass range, live out a main sequence followed by a supernova (SN) of some type~\citep{Woosley2015}.  The SNe considered in this work fall into three categories determined by the stellar mass of their progenitor: Type-II supernovae (SNe), hypernovae (HNe), and pair-instability supernovae (PISNe).  Depending on the mass of the star and mass of its host halo, the SN may completely disrupt the halo, ejecting the majority of gas and metal~\citep{Whalen2008b}, effectively preventing any further star formation until the gas has recycled back into the halo and cooled~\citep{Tumlinson2017}.  

Even if a \piii star is outside the mass range for SN, it will still emit ionizing radiation for its main sequence lifetime before collapsing to a BH. Such radiative feedback has been shown to shut down continuing star formation in its vicinity ~\citep{Whalen2008,wise2012b,Hopkins2019FB}, limiting the conversion of gas to stars in star forming regions.  Unfortunately, having the extremely high resolution ($\lesssim 20$ pccm\footnote{Comoving units have -cm appended to the base unit throughout this paper; the base unit is assumed to be proper.}, $M_{DM} \lesssim 10^4$ \msun) required to precisely model \piii SFF means that simulations of large (i.e., statistically relevant to the observable universe) volumes have not been able to approach modern redshifts~\citep{wise2012, xu2016, smith2015, Hopkins2019FB}.  To avoid the computational expense of precisely modelling the \piii era, some practitioners adopt a metallicity floor~\citep{Hopkins2018}, while others altogether disregard the effect of Pop III pre-enrichment on star formation~\citep{vogels2014}.  Neither of these simplifications account for the fact that enrichment by \piii stars is (a) non-uniformly distributed in space, (b) rare, and (c) necessary for enriched star formation.  Multiple recent works have highlighted the need for a more intelligent model for the \piii era: ~\citealt{jeon2017} and~\citealt{hicks2021} have shown that extremely low metallicity stars may form in halos that have been enriched by an external Pop III SN event, which represents a sequence of star formation that is impossible to model using the above simplifications. The lack of precision modelling of the \piii era is additionally cited as a potential cause for mismatches between simulations and observations of dwarf galaxies presented in \cite{wheeler2019}. \cite{regan2017} attempted to determine how super-massive black holes could form from primordial halos. They determine that the environment of a protogalaxy must avoid \piii star formation to enable the collapse of the protogalaxy into a black hole--simplistic models that assume a metallicity floor or ignore the metal contribution to enriched star formation will never be able to capture such a phenomenon.  These prior works and the limitations of currently used simplifications both highlight the need for a model that can bridge the gap between precision full-physics simulations and the current methods that ignore the ``initial conditions'' of metallicity in the universe. 

While artificial neural networks (ANN), and particularly deep convolutional neural networks (DCNN), have been used for image recognition for nearly a decade since AlexNet~\citep{alexnet} was used to classify images in the ImageNet dataset, they have also begun to foray into varied scientific applications, e.g., classifying observational images of radio galaxies~\citep{Aniyan2017}, predicting hydrodynamics quantities and modelling turbulence~\citep{Xiaowei2018, mohan2019}, and determining the halo occupation distribution of galaxies from a dark matter distribution~\citep{zhang2019}. Another class of emerging scientific ANN are emulators--models that supplant all or part of the application, of which, one generalized approach (DENSE) has been shown to accelerate some computations by up to $10^9\times$~\citep{kasim2020}.  The DENSE model shows great promise in scientific application, but is limited to scalar inputs; a model that evaluates the hydrodynamic state of a running simulation would require a different design paradigm. Motivated by the monumental advancements in the field, this work presents a novel surrogate model for pre-enrichment using trained DCNN to identify primoridal star forming regions.  This work will be coupled with a following work to predict \piii stellar feedback influence, where the entire model will function without resorting to halo finding and relax the severe resolution requirements of modelling the \piii era, accelerating the precision modelling of cosmological simulations. 

Precisely predicting the location of star formation is a two-fold task: a small 3D binary classifier based on seminal image classification architectures~\citep{He2015, huang2016, szegedy2014} will quickly analyze potential 3D regions; if one is classified as star forming, an inception based~\citep{szegedy2014} U-Net~\citep{ronneburger2015, ZHANG2020} adapted to 3D will predict star forming voxels.  The rest of this paper outlines the development of the first phase of this feedback algorithm, to identify star forming regions, as follows: In Section \ref{sec:sim_data} we outline the simulations and methods used to generate training data; Section \ref{sec:networks} presents the network architectures used and discusses the parameter exploration that led to our design; we test the DCNN designs and present the results in Section \ref{sec:results}, discuss the results in depth in Section \ref{sec:disc}, and summarize our findings in Section \ref{sec:conclusions}.

\section{Simulations and Data}
\label{sec:sim_data}
\begin{table*}
    \centering
    \caption{Source simulations for train/validation/test data.}
    \begin{tabular}{|c|c|c|c|c|c|c|r|}\hline
         Name      & SFF &$dx_{root}$[kpccm]   & $dx_{min}$ [pccm] & $L_{max}$ & $M_{DM}$[\msun] & $z_{final}$ & Cosmology\\\hline\hline
         PHX256-1  & w2012 & 10.183               & 19.89             &  9 & $2.384\times10^4$    & 13.57       & $\Omega_\lambda = 0.6889$ \\
         PHX256-2  & w2012 & 10.183               & 19.89             &  9 & $2.384\times10^4$    & 13.86       & $\Omega_m = 0.3111$\\ 
         P3N-128   & None  & 20.366               & 156.8             & 7  & $1.910\times10^5$    & 10.0        & $\sigma_8 = 0.811$\\
         PHX256-HYD & None &10.183               & 156.8             & 6  & $2.384\times10^4$   & 18.05       & $n = 0.968$\\
         \hline\hline
    \end{tabular}
    \tablecomments{ The width of the root grid cell, $dx_{root}$, width of a cell at the most refined level, $dx_{min}$, dark matter particle mass ($M_{DM})$ and maximum AMR level $L_{max}$ are shown for each simulation. All analysis in this work used data from these simulations up to the final redshift $z_{final}$. SFF denotes whether the simulation includes star formation (w2012 model from \citet{wise2012}) or not (None).  The final column shows the cosmological parameters common to all simulations \citep{Planck2014}.}
    \label{tab:test_sims}
\end{table*}
\label{sec:data_sims}
\subsection{PHX256 Simulations}
\label{sec:sims}
Training neural networks depends on copious amounts of data.  To that end, we used the astrophysical adaptive mesh refinement (AMR) simulation code \enzo \citep{enzo, Brummel-Smith2019} to produce two simulations (PHX256-1 and PHX256-2) from which to draw training, validation, and test data. Table \ref{tab:test_sims} summarizes the simulations used within this work. The PHX256-1,2 simulations are modeled after the Renaissance Simulations~\citep{xu2016} in terms of included physics, parameters, and resolution.   Both simulations have identical Planck 2014 cosmological parameters: $\Omega_\lambda = 0.6889, \Omega_m = 0.3111, \Omega_b = 0.04898, \sigma_8 = 0.811, n = 0.965$ \citep{Planck2014}, but use different random seeds in the initial conditions, generated using {\tt MUSIC}~\citep{hahn2011}.  While the Renaissance Simulations used a zoom-in simulation setup with a hierarchy of static nested grids, we simulate a periodic box of size comparable to the finest nested grid in the former, at identical mass and spatial resolution. The simulation volume for both PHX256 simulations is $2.61^3$ (Mpc)$^3$ in extent, with $256^3$ root grid cells and dark matter (DM) particles and 9 levels AMR; the cell width at the deepest AMR level is 19 pc.  With the given cosmology, the DM particles have mass $2.38\times 10^4$~\msun ($M_{dm}$), and the baryon mass ($M_b$) in a root grid cell of average density is $1.17\times10^3$~\msun.  

Refinement of the AMR grid occurs based on $M_b$ and $M_{dm}$, the baryonic and dark matter mass of a cell, respectively, with the minimum mass for refinement being 3 times the initial values above ($M_{min}$).  At each iteration, any grid cell on level $l$ with $M_{cell} \geq M_{min}\times 2^{-0.4l}$ will be refined to the next level; the grid refinement is super-lagrangian, so that the AMR levels will have $M_{cell} < M_{root~grid}$ . In addition to the mass criteria, any cells with \piii star particles are refined to the level such that the supernova radius parameter (10 pc) is resolved by $\geq$ 4 cell widths. With resolved halos defined as those with $\geq 100$ DM particles, the minimum mass of a resolved halo in the PHX256 simulations is $2.3\times 10^6$ M$_\odot$.  
We include radiation hydrodynamics using the Moray ray-tracing solver~\citep{wise2011}, with \piii and enriched star clusters as radiating sources. A uniform, $z$-dependent Lyman-Werner background is included to account for H$_2$ dissociating radiation sources originating from outside the simulation volume.  
Nonequilibrium primordial gas chemistry for the 9 species H, H$^+$, H$^-$, H$_2$, H$_2^+$, e$^-$, He, He$^+$, He$^{++}$ is computed, and radiative heating and cooling of the gas includes both primordial and metal-line cooling contrbutions as in \citealt{smith2008}. All simulations in this work were run on the TACC-Frontera supercomputer, with the PHX256-1,2 consuming $\sim$ 500~k cpu-hours total. 

Although the PHX256-1,2 simulations are distinct simulations, PHX256-2 uses identical physical models and resolution with different initial conditions as compared with PHX256-1.  The identical parameters include the total mass in the box, where the primary difference is in the spatial distribution of density peaks and troughs, whose magnitude is determined by the cosmological parameters common to both simulations. For this reason, we treat the two simulations as part of the same distribution of star formation examples. 

\subsection{Star Formation}
    Since this work aims to predict star formation, we will briefly review the relevant algorithms here.  For a more detailed review of star formation in \enzo, please refer to the \enzo documentation\footnote{\url{https://enzo.readthedocs.io/en/latest}}.  

    The PHX256 simulations both include \piii single star and \pii star cluster formation. At each grid timestep, the finest grid cell at each location is evaluated for star formation.  The \piii formation criteria and their parameter values which are checked are:
        \begin{itemize}
            \item Number density $n\geq 100$.
            \item H$_2$ density: $\rho_{H2}/\rho_b$ $\geq 10^{-3}$ .
            \item Metallicity\footnote{$Z$ denotes log metallicity relative to solar. With metal mass $M_z$, $Z = \log\frac{M_z}{M_b}-\log\frac{M_{z,\odot}}{M_\odot}$}: $Z\leq Z_c$ with $Z_c = -5.5$ for \piii formation. \pii formation requires $Z\geq Z_c$.
            \item The freefall time should be less than the cooling time: $t_{ff} < t_{cool}$
            \item Converging gas flow: $\nabla\cdot \boldsymbol{v}_{gas}< 0$.
        \end{itemize}
    If these criteria are met, a \pii ($Z>Z_c$) or \piii ($Z<Z_c$) star particle is formed from a sphere containing twice the mass of the star centered on the star forming grid cell.  In the \piii case, the particle represents a single star with mass taken from the a modified salpeter IMF of the form
        \begin{equation}
            f(\log M)dM = M^{-1.3}\text{exp}\bigg[-\bigg(\frac{M_{\mathrm{char}}}{M}\bigg)^{1.6}\bigg]dM,    
            \label{eqn:mass_fn}
        \end{equation} 
    with $M_{\mathrm{char}} = 20$ M$_\odot$.  \pii particles are formed if $Z > Z_c$ (the H$_2$ requirement is ignored). In this case, a single particle represents a radiating stellar cluster with $M_{min} = 1000$ M$_\odot$, assuming an unmodified Salpeter IMF.  Although the grid cell that was identified for \pii star creation must have $Z > Z_{crit}$, the gas surrounding likely has lower or higher metallicity: when mass-averaged into the star particle, the resulting particle may have $Z \neq Z_{cell}$.

\subsection{Data reduction and preparation}
\label{sec:data}
    Both simulations output the simulation state every 200 kyr, from $z=30$ to the final redshift noted in Table~\ref{tab:test_sims}. Between $z=30$ to $z=10$, there will be 1806 individual data outputs that are a snapshot of the entire simulation domain. This project serves as a proof of concept, so uses all simulation outputs up to their most progressed state ($z_{final}$ in Table \ref{tab:test_sims}).  Future work will progress simulations to $z \leq 10$ to incorporate lower-$z$ data into the datasets. In this method, we assume that \piii formation is independent of $z$.  In other words, we train our models on star formation events across a range of redshifts without treating data at different redshifts as an isolated distribution of star formation. The justification for this is the set of criteria that must be met to form a Pop III star in a fully resolved simulation (bulleted list above) depends only on local conditions which are decoupled from global conditions. The one exception is the globally evolving UV Lyman-Werner background, which affects the H$_2$ fraction of the gas, which in turn enables the gas to cool and condense into stars. However, since our model includes the H$_2$ fraction, this global influence is taken into account. In support of our assertion that redshift is not an important parameter of the problem is the analysis of \cite{xu2016}, who showed that the statistical properties of high-z dwarf galaxies in the Renaissance Simulations were insensitive to redshift, but rather principally dependent on the halo virial mass. There, galaxies form from gas pre-enriched by \piii stars whose formation process is modeled directly. 
    
    The simulation outputs occupy $>75$ TB in their unprocessed state at the current redshifts. Although copious, this raw data is not acceptable for input to a neural network, so the both simulations are post-processed to generate training, validation, and testing data, as well as to reduce the size of the data to a more manageable footprint.  The final dataset counts, as derived from the following process, are tabulated in Table~\ref{tab:datasets}.
    
    To generate model input data, we use pairs of snapshots from a single simulation, $\{D, D_{-1}\}$, where $D$ is the current output and $D_{-1}$ is the output immediately prior. Each $D$ output is checked for new star particles since $D_{-1}$.  If found, we generate a uniform grid with volume (9.98 kpccm)$^3$ centered on the new star particle in $D$ using {\tt YT}~\citep{turk2011}.  Each hydrodynamic and color field in the region in $D_{-1}$ is saved to this cube.  To label the star location, we flag cells of a $3^3$ cube centered on the star forming grid cell as star forming voxels.  For sample augmentation, we additionally generate $n_{\text{shifted}}$ volumes that are centered randomly, but still contain the target star particle.  After all new stars are accounted for, we generate random samples of regions in $D_{-1}$ with the restriction that any candidate volume must have a volume average density ($\langle \rho\rangle$) greater than the mean($\bar\rho$): $\langle \rho\rangle/\bar \rho \geq 1.0$.  There are $n_{\text{star}}(1+n_{\text{shifted}})+n_{\text{star}}\times 50$ samples generated for each snapshot of the simulation, where we set $n_{\text{shifted}} = 5$. 
    
    For PHX256-1, samples are separated into training and validation, which is randomly determined for each volume during initial data reduction while ensuring that volumes with the same star particle belong to the same split. This proof of concept uses  81,512 (9380 with stars) training samples with 6,439 (412 with stars) validation samples. There is a lower ratio of star containing volumes in validation because we take care to remove volumes that also contain a star in the training dataset. To  ensure  a  pristine  test dataset, we perform data reduction as above on PHX256-2, generating 27,231 (3,782 with stars) samples exclusively for the testing dataset.
    
    In the final step of preparation, each sample is scaled by the standard deviation and mean of all training data, i.e., for density, $\rho_{scaled} = (\rho - \langle \rho\rangle) / \langle \sigma(\rho)\rangle$ for $\langle \rho\rangle$, the average voxel density in training and $\langle \sigma(\rho)\rangle$, the averaged standard deviation of voxel density in training data.

\begin{table}
    \centering
    \caption{Summary of final dataset for training, testing, and validation}
    \begin{tabular}{|c|c|c|c|c|}\hline
         Partition  & $N_{stars}$ & $N_{SFR}$ & $N_{nSFR}$ & $N_{total}$\\\hline\hline
         Train      &   1564    & 9380      & 72132 &   81512\\
         Validation &   67      & 412       & 6027  &   6439\\
         Test       &   630     & 3782      & 23449 &   27231\\ 
         \hline\hline
    \end{tabular}
    \tablecomments{The raw number of stars in each partition is given by $N_{stars}$, the number of star forming regions after data augmentation by $N_{SFR}$, and the total number of non-star forming regions by $N_{nSFR}$}
    \label{tab:datasets}
\end{table}
    We choose which hydrodynamic fields to train on based on our knowledge of star formation criteria in \enzo.  Each volume has 5 channels as input: baryon density ($\rho_b$), H$_2$ density ($\rho_{H2}$), gas velocity divergence ($\nabla\cdot \nu$), total metallicity ($Z_{sum}$), and total (kinetic + thermal) energy ($E_T$).  Our selection of fields can also be based on physical intuition: \piii star formation occurs only in regions with high $\rho_B$, $\rho_{H2}$, and very low $Z_{sum}$.  $E_T$ serves as a dual probe: strong radiation fields will increase temperature, thereby increasing thermal energy, while fast-moving gas would increase the kinetic energy.  Star formation is not expected in either fast moving or hot gas, so having a high value of $E_T$ should disqualify the region for \piii star formation.  The effect of $Z_{sum}$ is essentially binary: $Z_{sum}\leq Z_{crit}$ should enable \piii star formation, but any other case should immediately disqualify the region. The $\nabla\cdot\nu$ field should immediately disqualify star formation if the gas is not converging.  
    
    Although a model with fewer fields may appear effective, losing any of these probes into the hydrodynamical state would be expected to not be as robust across all redshifts. For example, if $\rho_{H2}$ is ignored, the model may start to fail after the LW background becomes strong enough to dissociate H$_2$ in relatively dense gas.  If $E_T$ were removed, the model would have to learn to infer the energy of the gas from $\nabla\cdot\nu$, while losing all probes into the temperature, except that $\rho_{H2}$ and $\rho$ would likely be lower at higher temperature.

\section{Network Design}
\label{sec:networks}
%
%
\begin{figure*}
    \centering
    \includegraphics[width=0.7\textwidth]{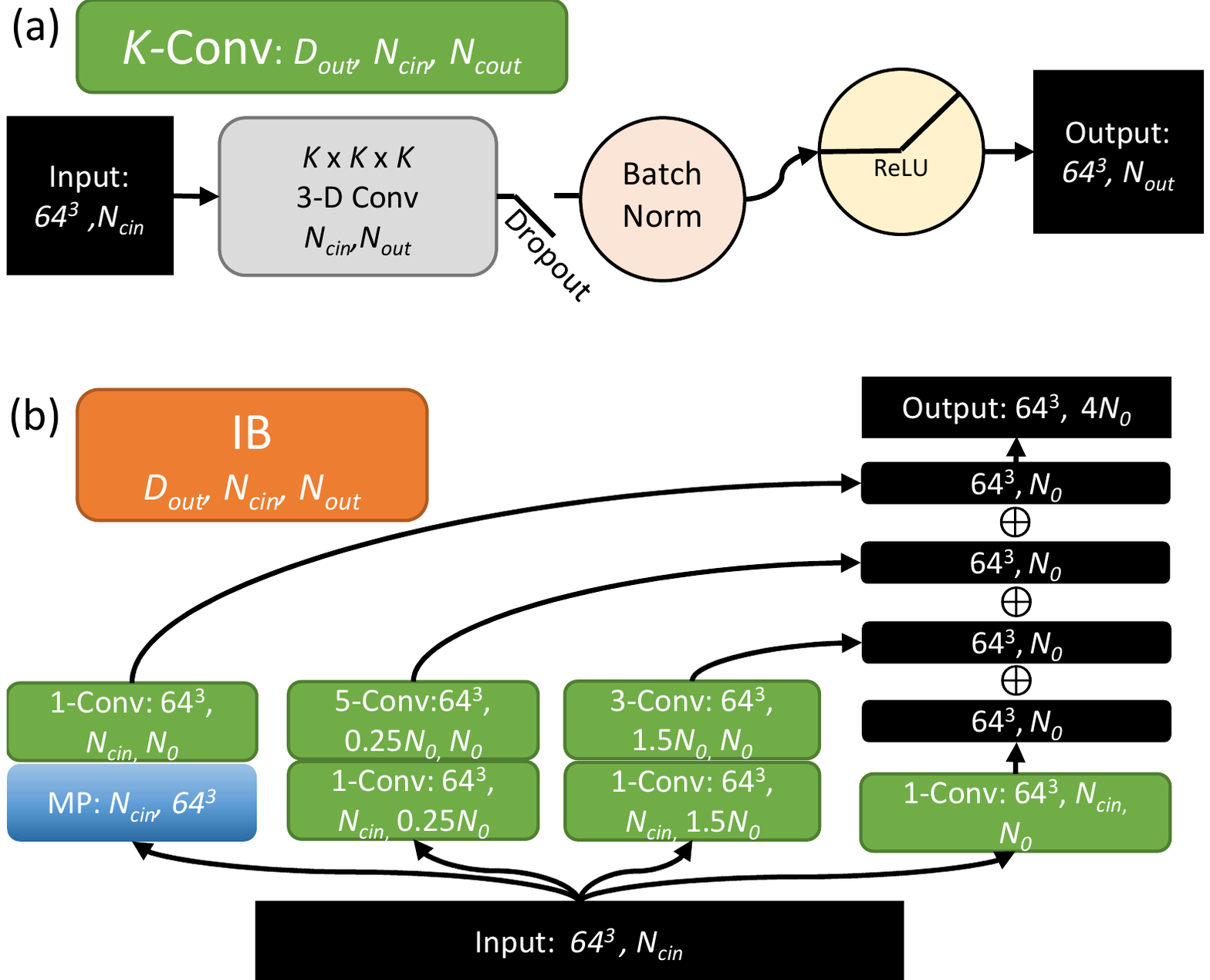}
    \caption{The basic convolution block and inception block used in this work.  (a): 3D convolution with $K^3$ filter ($K$-Conv) used throughout this work.  By default, every convolution is followed by dropout, batch normalization, and ReLU activation.  The output dimension is unchanged, with $N_{out}$ channels.  (b): An example Inception block (IB) used throughout this work.  A copy of the input is processed in four branches with the outputs from each branch concatenated ($\oplus$) together to form the final output.  The convolutions follow the convention of $K$-Conv in (a), except that the 1-Conv blocks in the middle two branches have no activation function and dropout is applied only after the last convolution in each branch. Network width can be tuned by the parameter $N_0$.  The channel numbers in the IB are used in SINet (section \ref{sec:networks-s1}), IBs in IUNet (section~\ref{sec:networks-s2}) use simplified values with each branch outputting $N_0/4$ channels.}
    \label{fig:conv}
\end{figure*}
%
%
In a grid-based simulation code like \enzo, the state of the simulation is stored on a fixed grid, where quantities like baryon density are stored as cell-centered quantities. Every quantity that is advanced by the simulation is tracked on the grid, or derived from quantities that are tracked there.  Therefore, when inspecting a hydrodynamical simulation, it can be viewed as a 3-dimensional volumetric image, where each hydrodynamic quantity is analogous to the RGB color channels of a typical image. Of course, different hydrodynamic fields may carry nearly independent information, so the analogy to color channels is only surface deep; nonetheless, this logic leads us to classify volumes of a simulation as if they were volumetric images with hydrodynamic information as channels.  Such an approach lets us take advantage of the numerous developments in the field of computer vision for this problem.  

We use a two-stage approach to predict localized \piii star formation. The first stage ($S1$) is a classifier used to quickly decide if a potential region is capable of forming stars in any of its volume.  If the classifier identifies a star forming region, it is followed by a more complicated voxel segmentation network ($S2$), used to identify which voxels within the volume are forming stars.  The module composed of $S1$ and $S2$ (\sss) must agree on the star forming state of the volume in order to classify the volume as star forming.  All of the architectures presented here are implemented in Pytorch~\citep{pytorch} and were trained using 4 K80 GPUs on the SDSC-Comet supercomputer. 
%
%
\begin{figure}[t]
    \centering
    \includegraphics[width=0.49\textwidth, clip, trim=1.8cm 0cm 1.8cm 0cm]{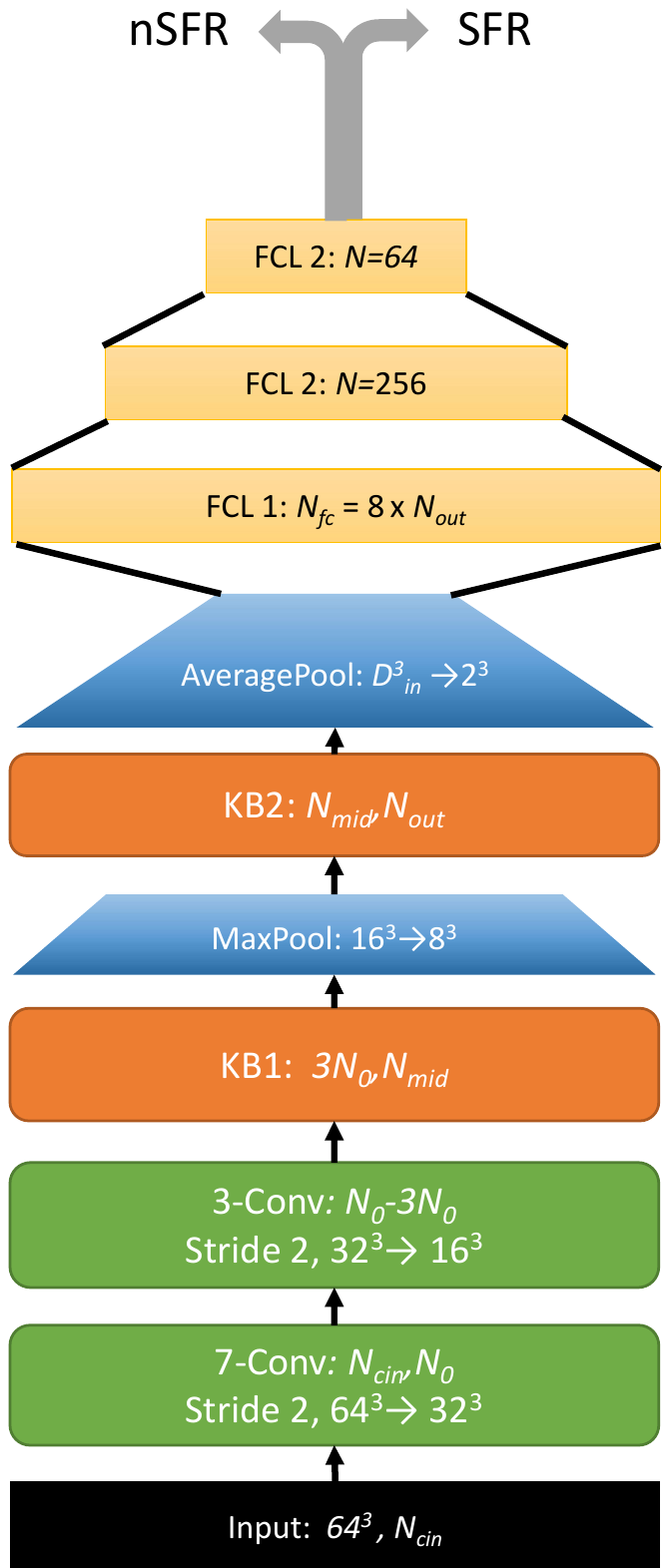}
    \caption{Small classifier design used in this work.  Blocks indicate a layer of processing, with input and output channels notated. Data is initially processed in two standard convolutions with filter $7^3$ and $3^3$, with stride 2 to reduce dimension by half.  These are followed by two key blocks(KB1, 2), with another max-pooling after KB1.  Before the fully connected layers (FCL), the output of KB2 is average pooled to $\{N_{batch}, N_{out} , 2,2,2\}$.  All layers use ReLU activation.  The final output is two classes, \{nSFR, SFR\}.  The parameter $N_0$ can be used to tune the width of the network while maintaining the same architecture.  In principle, the KB can any specialized processing block, e.g., an inception, densely connected, or residual block, as seen in Figures~\ref{fig:conv} and \ref{fig:keyblocks}.}
    \label{fig:smallincep}
\end{figure}
\begin{figure}[t]
    \centering
    \includegraphics[width=0.49\textwidth, clip, trim=3cm 3cm 3cm 2cm]{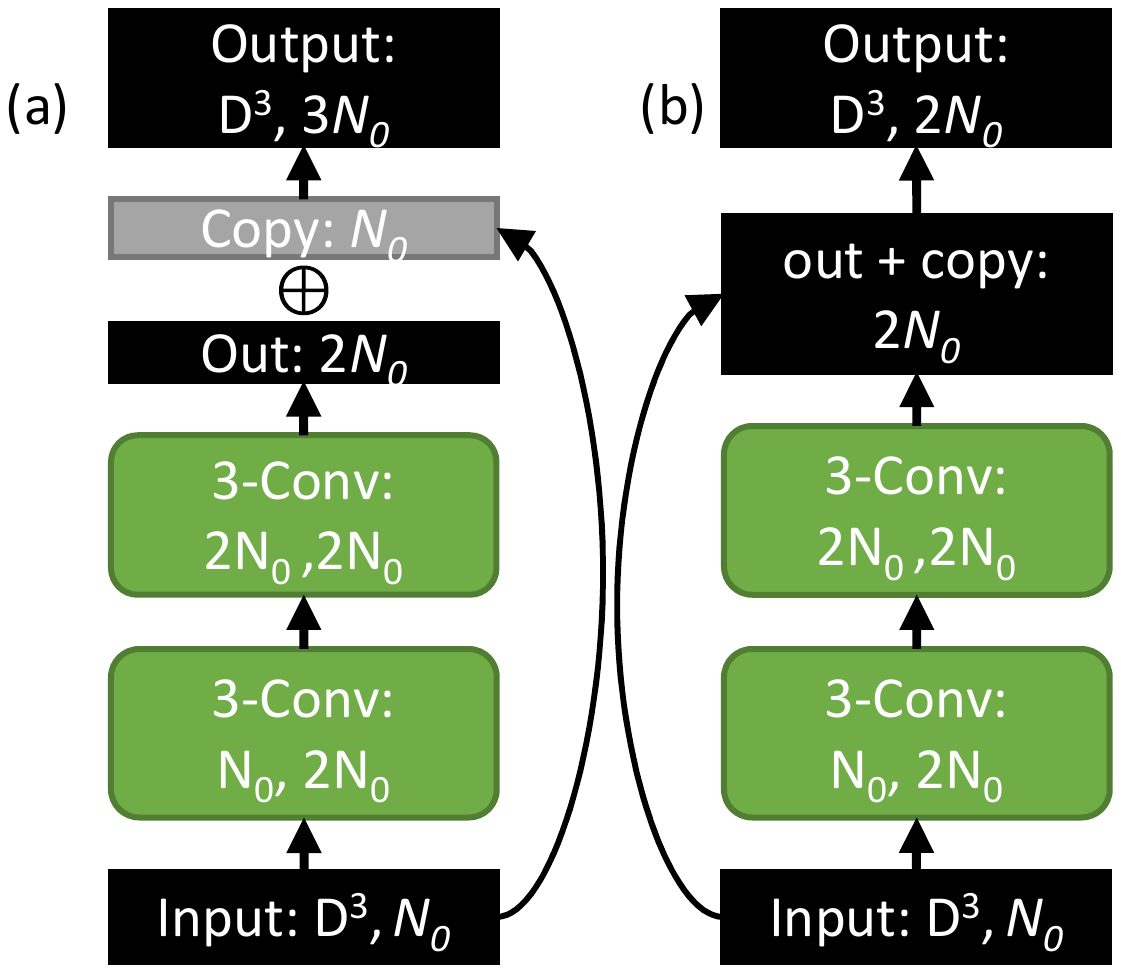}
    \caption{Key blocks for small classifier. Both versions accept a $D^3$ dimensional input with $N_0$ channels. (a): The DenseNet block that replaces KBs in SDNet.  The input is copied, and concatenated ($\oplus$) to the output after the convolutions.  (b): The ResNet block that replaces KBs in SRNet.  The input is copied and added to the output after the convolutions.}
    \label{fig:keyblocks}
\end{figure}
%
%
\subsection{Stage 1: Classification Network}
\label{sec:networks-s1}

    We tested various network architechtures to be used for classifying regions as star forming (SFR) or not star forming (nSFR).  Initial tests included 3D adapted versions of 16-layer ResNet~\citep{He2015}, 16-layer DenseNet~\citep{huang2016}, and GoogLeNet as described in \cite{szegedy2014}, however it quickly became apparent that these models were too complex for the task at hand.  To reduce the model complexity, this work uses small classifiers based on ideas from those seminal papers, with vastly reduced network depth.  In addition to changing the depth of the network, this work also implements several changes to the above architecture designs:
    \begin{itemize}
        \item Number of input channels represents input hydrodynamic fields and is a hyperparameter.
        \item The network width is another hyperparameter, tunable using the number of channels at the first layer, $N_0$.
        \item Each convolution, regardless of base architecture, is followed by dropout, batch normalization, and ReLU activation, with exceptions in Inception blocks (IBs) (Figure~\ref{fig:conv}).
    \end{itemize}

Figure \ref{fig:conv} describes the basic convolution and IBs used throughout this work.  By default, the convolution block has a 3D $K^3$ convolutional filter, dropout with probability 0.2, batch normalization~\citep{ioffe2015}, and ReLU activation. The default stride and padding are set such that the input dimension is unchanged. In the interest of preserving spatial relationships while reducing overfitting, dropout zeros the output of an entire channel in the output of the layer~\citep{tompson2014}. The IBs presented in Figure~\ref{fig:conv} are used in the classifier architecture.  In principle, all channel numbers in the IB are tunable hyperparameters, however this parameter space was not explored in this work.  

To create a modular and minimal effective classifier architecture, we designed the small classifier networks as shown in Figure \ref{fig:smallincep}.  These architectures are inspired by GoogLeNet, DenseNet, and ResNet, but are significantly shallower.  After initial convolutions common to each architecture, each small classifier network is composed of only two key blocks, as seen in Figure~\ref{fig:smallincep}.  The key blocks may be inception modules (SINet), standard convolutions with residual skips (SRNet), or densely connected convolutions(SDNet)(see Figure~\ref{fig:keyblocks}).  The rest of this work will focus exclusively on the use of small SINet, SDNet, and SRNet, as there was no appreciable gain in accuracy to using the full architectures from ResNet, DenseNet, or GoogLeNet, while inference and training were much slower with significantly higher memory consumption.  
    
Before final classification, we remove dependence on input dimension by reducing to a $\{N_{channel}, 2, 2, 2\}$ volume via average pooling before the fully-connected layers.  All network variations use Adam optimization~\citep{adam} with $\beta_1 = 0.9$ and $\beta_2 = 0.999$. Although this is a physics oriented problem, the prediction is a simple binary classification.  With this in mind, we minimize the cross entropy loss given by

\begin{equation}
    L(\hat y, y, c) = \bigg\{w(c)\bigg[-y_c\hat y_c+\log\bigg(\sum_j\exp(\hat y_j)\bigg)\bigg]\bigg\},
    \label{eqn:xent}
\end{equation}
\begin{equation}
        L = \frac{\sum_c L(\hat y, y, c) }{\sum_c w(c)}
\end{equation}
where $w(c)$ represents weights given to the $c^{th}$ class (here, classes = $\{0,1\} = \{$nSFR$, $ SFR$\})$, the network output ($\hat y_j$) and true label ($y_j$).  

A uniformly sampled dataset of the simulations would have $\lll 0.001\%$ of the volume classified as SFR.  We deal with this extreme bias via multiple methods: a) sampling is selective, as described in Section \ref{sec:data}, and b) we use weights in the loss function with $w = [1, 4]$ for \{nSFR, SFR\}. Several combinations of weights were explored via random search in hyperparameter tuning, where we chose the weight value to minimize false positive results while maximizing accuracy (i.e., we prefer false negative predictions over false positive predictions).  We additionally employ $L_2$ regularization to guard against overfitting with $L_2$ parameter $\lambda = 10^{-4}$.  We use a plateau method to reduce the learning rate: the rate is reduced by half if the loss on the validation dataset has not reduced for 10 epochs until a minimum learning rate of $10^{-8}$.  We also employ a checkpointing method, saving model weights and parameters each time a new record low loss is achieved on validation data.  The final model used in the following sections is the checkpoint using these best-loss weights. 

\begin{table}[ht]
    \centering
    \caption{Tested classifier architectures for S1.}
    \begin{tabular}{|c|c|c|c|c|c|c|c|}\hline
        Model       & Key Block &  $N_0$ & $N_{tp}$[M]   & $N_{b}$  & $E_{BL}$    \\\hline\hline
        SINet       & IB        & 16        &  0.41         & 480    & 160      \\
        SDNet       & Dense     & 32        &  14.67        & 460    & 156      \\
        SRNet       & Residual  & 32        &  21.29        & 460    & 193      \\

        \hline\hline
    \end{tabular}
      \tablecomments{All models were trained on input data with dimension dim$^3$=$64^3$.  $N_{tp}$ is the number of trainable parameters in the model, and $N_b$ is the batch size. The Key Block describes the architecture used in place of KB1 and KB2 in Figure~\ref{fig:smallincep}.  The epoch of the checkpoint with the best validation loss, which is used for evaluation in this work, is recorded in $E_{BL}$. } 
    \label{tab:classifiers}
\end{table}

\subsection{Stage 2: Segmentation Network}
\label{sec:networks-s2}
%
%
\begin{figure*}[t]
    \centering
    \includegraphics[width=0.8\textwidth,clip, trim=0.0cm 3.9cm 0.0cm 3.9cm]{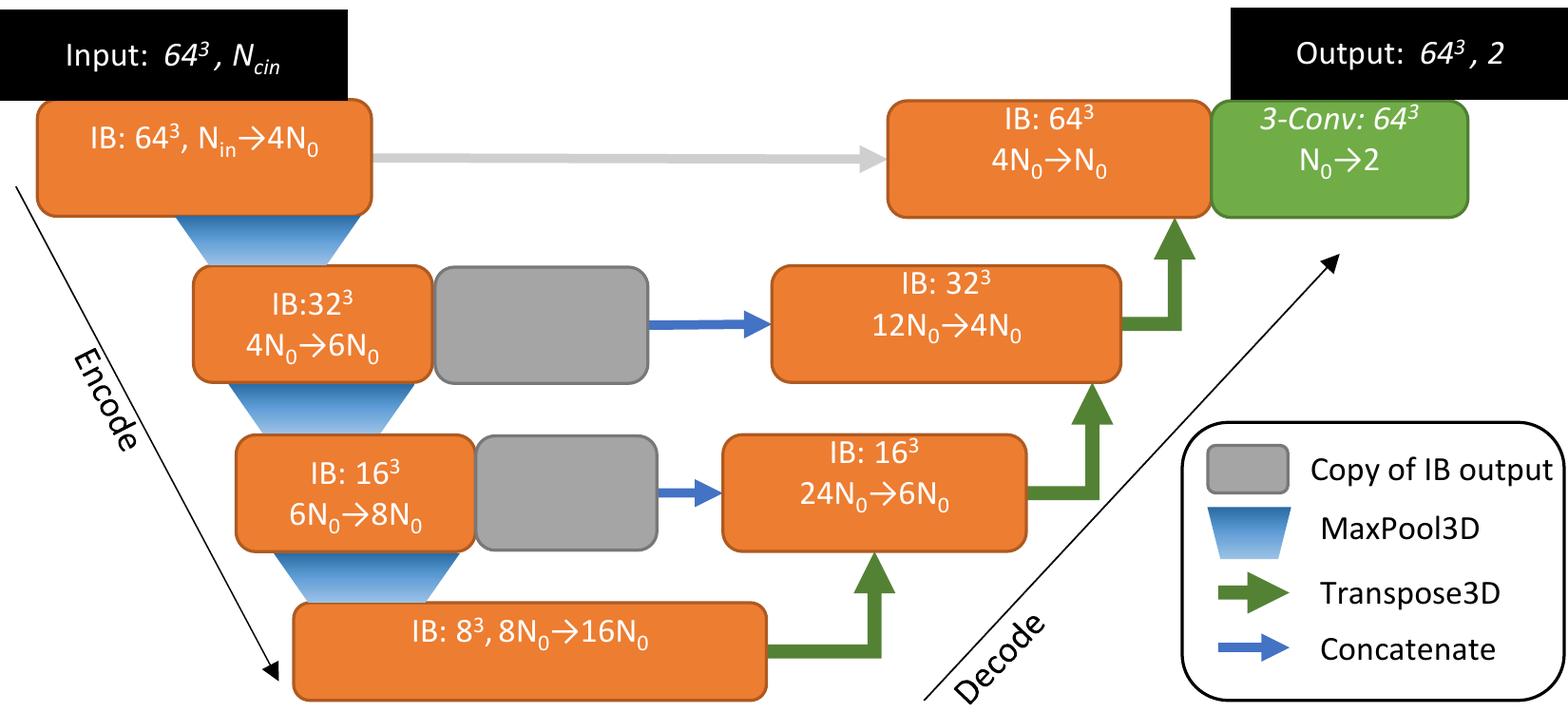}
    \caption{Customized Inception U-net used throughout this work.  The top concatenation arrow (grey) is not used in this work to force processing at deeper layers of the network.  The parameter $N_0$ is used to determine the width of the network as a whole, where we use $N_0 = 4$.  The final convolution takes the network output of $N_0$ channels to 2 classes, \{nSFV, SFV\}. Each orange block is an IB (Figure~\ref{fig:conv}) annotated with data dimension, input channel number and output channel number. The output of the encoder branch IBs is copied and concatenated to the appropriate level of the decoder branch before the IB is processed. The MaxPool3D operations use filter 3, stride 2, padding 1, so that the output has dimension reduced by half.  The Transpose3D blocks on the right side have kernel size 2, stride 2, and padding 0 to mirror the reduction of the max pooling in the encoder branch.  Each IB here follows the construction presented in Figure~\ref{fig:conv}.}
    \label{fig:iunet}
\end{figure*}
%
%
After classification, we use deep neural network architectures designed for pixel segmentation, which we have adapted to 3D voxel segmentation, to identify star forming voxels (SFVs) in stage 2 (S2).  Two architechtures were tested here: U-Net \citep{ronneburger2015}, and a variation of U-Net motivated by \cite{ZHANG2020}  that uses IBs instead of standard convolutions (IUNet).  In early testing, U-net was plagued by false positives, and was abandoned in favor of the IUNet architecture presented in Figure \ref{fig:iunet}.  As outlined in section~\ref{sec:data}, each star in a SFR is labelled by 27 cells in a $64^3$ box, so the data is extremely biased. To deal with this bias, we use class weights in the cross entropy loss function, with $w = [1,9]$ for classes \{SFV, non-SFV(nSFV)\}. As with $S1$, this value is the result of minimizing false positive results, while maximizing volumetric accuracy during a random search centered at $w = [1,8]$ as used in \citet{zhang2019}.  As with the classifier networks, we use the Adam optimizer with $L_2$ regularization using $\lambda = 10^{-5}$. We use an initial learning rate of $5\times10^{-3}$ with the same learning rate plateau adaptation method as S1.

IUNet functions similarly to an autoencoder, replacing fully connected layers with convolutional IB.  The encoding branch reduces dimensionality via max pooling while increasing the number of channels via concatenation and convolution operations until the lower bottleneck, where the process is reversed in the decoding branch. The encode and decode layers are connected by concatenating the encode output to the decode input at a given layer.  Finally, the output is reduced to two channels, representing nSFV and SFV, for each voxel in the region. 

The skips in IUNet allow more efficient back-propagation and allow information to flow directly from encoding branch to decoding branch, potentially reducing the size of the training set required to attain a robust and generalizable model~\citep{ronneburger2015}.  The skip connections may also have a detrimental effect here though, as star formation is highly correlated to peaks in the density field, so the entire network may be skipped with predictions being made directly from the input fields.  As seen in \cite{zhang2019}, to reduce the direct communication from the input fields to the final prediction, we removed the top most skip connection to force processing at deeper layers of the network.  The IB architecture has several different sizes of convolutional filters ($K = 1^3, 3^3, 5^3$), which can give sensitivity to different scales of features in the data, e.g., 100 pc scale infalling gas toward a density peak 1-10 pc in radius.  We hypothesize that this multi-scale sensitivity from IBs is likely why IUNet outperformed our standard U-net architechture early on.  

Although $S2$ minimizes cross-entropy loss for training, we also calculate the intersection over union ($IoU$) metric to judge the quality of the $S2$ voxel-wise predictions. If $P_t$ is the set of positive predicted voxels, and $P_{T}$ is the set of ground truth (GT) positive voxels, the $IoU$ is given by
\begin{equation}
    IoU = 1-\frac{P_t\cap P_T}{P_t\cup P_T}.
    \label{eqn:iou}
\end{equation}
 With this representation, $IoU = 0$ is a pefect prediction with no false negatives or false positives.


\section{Results}
\label{sec:results}

\subsection{S1 Volume Classifier}
\label{sec:results-s1}

\begin{figure*}
    \centering
    \subfloat[Training]{\includegraphics[width=0.48\textwidth]{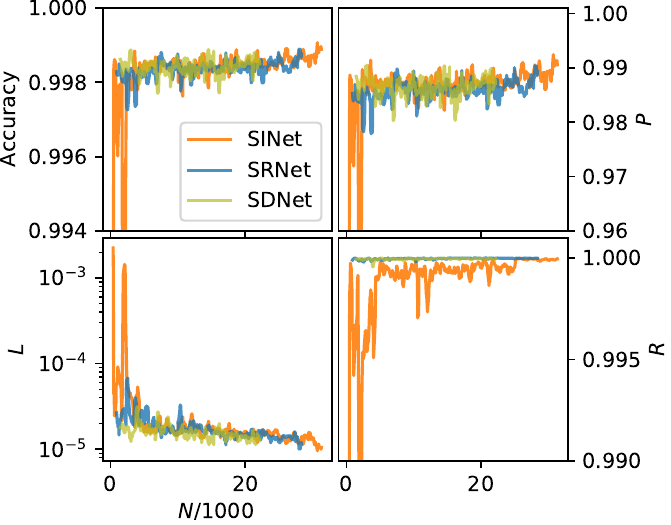}

        \label{fig:s1-training}
    }
    \subfloat[Validation]{\includegraphics[width=0.48\textwidth]{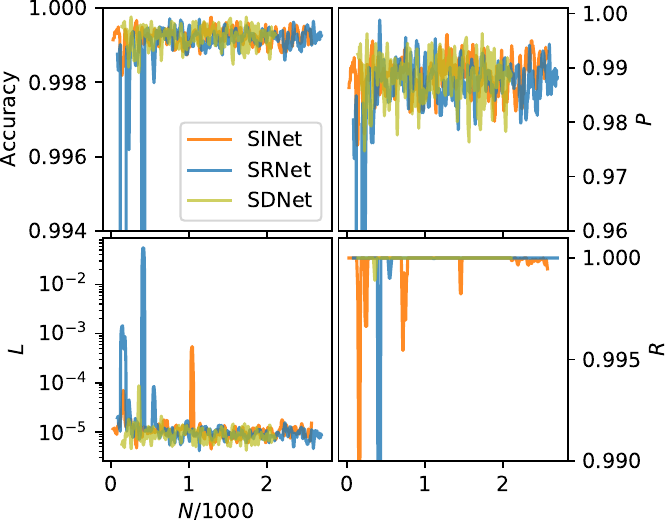}
        \label{fig:s1-validation}}

    \caption{Training (a) and validation (b) results for the training of small classifier architectures.  We further quantify the accuracy with the precision ($P$) and recall ($R$) measures, as defined in section~\ref{sec:results-s1}.  Despite the bias in our samples, the errors in S1 are dominated by false positives indicated by the relatively lower $P$ across all architectures. $N/1000$ represents one output every 10 iterations within each epoch--validation has fewer data points because there are fewer validation iterations in each epoch.}
    \label{fig:s1results}
\end{figure*}

We use several primary metrics to quantify the success of an $S1$ model: (a) Raw accuracy as simply $A = N_{correct}/N_{total}$, (b) precision, $P = P_T/(P_T+P_{f})$, with true positives $P_T$ and false positives $P_f$, and (c) recall, $R = P_T/(P_T+N_f)$ with false negatives $N_f$.  We will also extend the $P$ and $R$ metrics with the $F_1$ score, given by $F_1 = P_T/(P_T+0.5(P_f+N_f) = 2\times P\times R/ (P + R)$ to provide another measure of classification ability. With these definitions, perfect result would have $P=R=F_1=A=1$. We finally present the completeness~\citep{rosenberg2007} considering SFR and nSFR as two clusters that must be separated by the $S1$ classifier. Although $A$ is of limited use in these biased datasets, we present it as evidence that the models far outperform blanket SFR or nSFR predictions. Figure~\ref{fig:s1results} presents the $P$ and $R$ measures along with the loss for training and validation of all tested $S1$ architectures.  
\begin{table}
    \centering
    \caption{Results of $S1$ on testing data from PHX256-2. }
    \begin{tabular}{|c|c|c|c|c|}\hline
         Model      & Accuracy              & $P$       & $R$       & $F_1$  \\\hline\hline
        SINet     & 27201/27231 (0.9989)   & 0.9952    & 0.9947  & 0.9950  \\
        SDNet     & 27201/27231 (0.9989)   & 0.9927    & 0.9976  & 0.9951 \\
        SRNet     & 27198/27231 (0.9988)    & 0.9924    & 0.9968  & 0.9946 \\
        IUNet     & 27112/27231 (0.9956)    & 0.9932    & 0.9960  & 0.9946 \\\hline\hline
    \end{tabular}
    \tablecomments{We also present $S2$ using IUNet as if it were used to classify SFR. Every classifier tested is able to achieve high accuracy $\gtrsim 99.8\%$, reinforced by very high precision $P$, recall $R$, and $F_1$ score (as defined in Section \ref{sec:results-s1}). }
    \label{tab:s1results}
\end{table}
SDNet and SRNet, despite having similar memory requirements as SINet in training, have significantly more trainable parameters.  Their increased complexity appears to affect how quickly they converge to a trained state, as they seem to have similar accuracy to SINet at much earlier epochs.  This effect is particularly noticeable in the recall.  On testing data results presented in Table \ref{tab:s1results}, we find that all three S1 models perform extremely well, with all three having $A \geq 0.9988$.  All three additionally have very similar precision and recall, as measured on their volume-wise classification.  In $F_1$ score, SDNet performs best ($F_1 = 0.9951$), followed closely by SINet ($F_1 = 0.9950$) and SRNet ($F_1 = 0.9946$). We do not measure the inference rate of $S1$ as it does not produce a desirable prediction in this work; it needs to be coupled with $S2$ to localize star formation to a precise region.
%
%

\subsection{S2 Voxel Segmentation}
\label{sec:results-s2}
\begin{figure}
    \centering
    \includegraphics[width=0.46\textwidth]{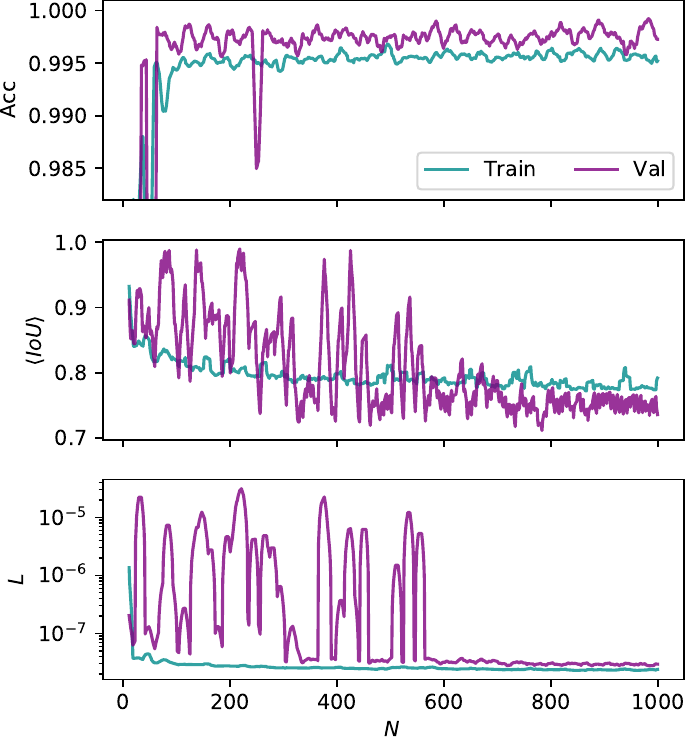}
    \caption{Accuracy (top, $Acc$), averaged $IoU$ (middle, $\langle IoU \rangle$), and cross entropy loss (lower, $L$) from training and validation of IUNet. $N$ represents intra-epoch recordings, not end of epoch data, and validation data has been expanded to align with training. These data reflect epochs of training up to the best recorded validation loss.}
    \label{fig:s2_trainval}
\end{figure}

Loss, $\langle IoU\rangle$, and volumetric accuracy (Acc) for training and validation for $S2$ are presented in Figure~\ref{fig:s2_trainval}. $S2$ is a capable volume classification model, with $A > 0.994$ in both training and validation. After training, $S2$ was also applied independently to the test dataset (Table~\ref{tab:s1results}), where $S2$ classified regions with $A = 0.9956$ with $P = 0.9932$, $R=0.9960$, and $F_1 = 0.9946$, indicating that $S2$ is as capable as the $S1$ models at volumetric classification for SFRs.  Despite this high accuracy in classifying regions, $S2$ struggles to match individual star forming voxels, with $\langle IoU\rangle >0.819 $.  Indeed, if the voxel-wise accuracy is quantified by averaged precision and recall, we find $\langle P\rangle \geq 0.258$ and $\langle R\rangle \geq 0.875$ yeilding $\langle F_1\rangle \geq 0.399$.  These combined measures show a significant propensity to false positive SFV.  This is reinforced by simple statistics; there are an average of 36.34 ($\sim3.3^3$ voxels) true SFV in each SFR, with $\sigma = 14.78^[$\footnote{$\sigma$ here represents the standard deviation of the distribution}$^]$ while $S2$ predictions average 89.86 ($\sim4.4^3$) SFV per SFR with $\sigma = 45.9$. 
\begin{table*}
\centering
\caption{Results of different configurations of \sss modules on testing data from PHX256-2. }
    \begin{tabular}{|c|c|c|c|c|c|c|c|}\hline
            $S1$   & $\langle IoU\rangle$    & $P$        & $R$    & $F_1$ & Completeness         & TSS  & HSS\\\hline\hline
         SINet     & 0.8191                 & 0.2590     & 0.8750 & 0.9950(0.3997)    &  0.9752 (0.2354) &  0.9940 (0.8571) & 0.9942(0.3626) \\
         SDNet     & 0.8196                 & 0.2584     & 0.8759 & 0.9954(0.3991)    &  0.9746 (0.2349) &  0.9964(0.8571) & 0.9943(0.3626)  \\
         SRNet     & 0.8195                 & 0.2592     & 0.8754 & 0.9950(0.3999)    &  0.9724 (0.2356) &   0.9956(0.8571)& 0.9937(0.3626)    \\
         \hline\hline
    \end{tabular}
    \tablecomments{Each row represents the \sss module using a different architecture for $S1$.  The averaged $IoU$ ($\langle IoU\rangle$), $P$, and $R$ pertain to voxel-wise predictions within the regions.  $F_1$, Completeness, true skill score (TSS), and Heidke skill scores (HSS) are reported in volumetric classification and voxel-wise classification in parenthesis.}
    \label{tab:s3results}

\end{table*}

The predicted region still covers the GT voxels, as indicated by the high value of the voxel-wise recall.  Aside from recall, this behaviour can be quantified in a spatial sense: if we define a star forming center as the average location of SFV, the euclidean distance from the prediction center to GT center is another measure of prediction quality.  Here, we measure the euclidean distance in voxel widths, and find that the average distance between SFV center and GT center is 1.90 voxel-widths.  Further, the standard deviation of the distribution of distances is 2.368 voxel-widths, while the largest observed distance is 23.72 voxel-widths.  In examining the distribution of distances, we find that 98.1\% of SFV centers are identified within 10 voxels of the GT central point, and that 0.796\% of centers are more than 15 voxel-widths from the ground-truth center.  This metric will likely be improved with further training and inclusion of more data.

The single largest shortcoming of using $S2$ for all processing is in its inherently slower computation time.  Processing every volume through $S2$ can only be done at a rate $R_{proc}$ of 54.67 volumes/second, when only timing the inference of the model.  Given that $> 90\%$ of volumes contain no stars, a quick, simple model that can discard obvious nSFR would greatly expedite the inference of the final model.  For this reason, we chain together $S1$ and $S2$ to form the \sss module.

%
%

\subsection{\sss Module}
To process a sample through the \sss module, we filter the sample volumes using $S1$, only passing those where $S1$ predicts SFR on to $S2$.  The two models are designed to be independent to test the efficacy of different combinations of architectures.  The results of this test are tabulated in Table~\ref{tab:s3results}.  The completeness of volumetric classification is $0.9724$, while voxel-wise completeness is lower, at $0.2349$.  This lower classification skill in voxel segmentation is obvious in all presented metrics including true skill score (TSS) and Heidke skill score (HSS). The low voxel-wise $P$ and high $R$ imply that all \sss variations suffer from false positive voxel predictions, reinforced by relatively lower $F_1$ and voxel-wise skill metrics.  $S2$ has high volumetric classification skill overshadowed by slow processing; implementing an $S1$ model to filter regions increases the rate of processing samples ($R_{proc}$), by up to 8$\times$, and may increase some or all of the $F_1$, $P$, and $R$ scores for IUNet seen in Table~\ref{tab:s1results}.  Also of note is that the actual architecture of $S1$ is largely irrelevant given the models developed here; SINet, SDNet, and SRNet all perform very similarly in all evaluated metrics.  Unfortunately, the inclusion of $S1$ was unable to significantly improve the voxel-wise predictions (e.g., $\langle IoU\rangle$ or completeness) of $S2$ by prefiltering nSFR--this implies that there are very few false positive regions being identified by $S2$ that are not also classified as SFR by the various $S1$ models.

\begin{figure*}
    \centering
    \subfloat[Train]{
        \includegraphics[width=0.48\textwidth]{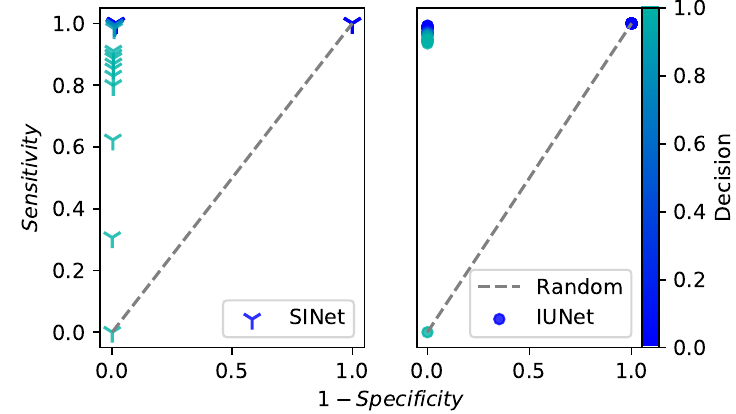}
        \label{fig:roc_train}
        }
    \subfloat[Test]{
        \includegraphics[width=0.48\textwidth]{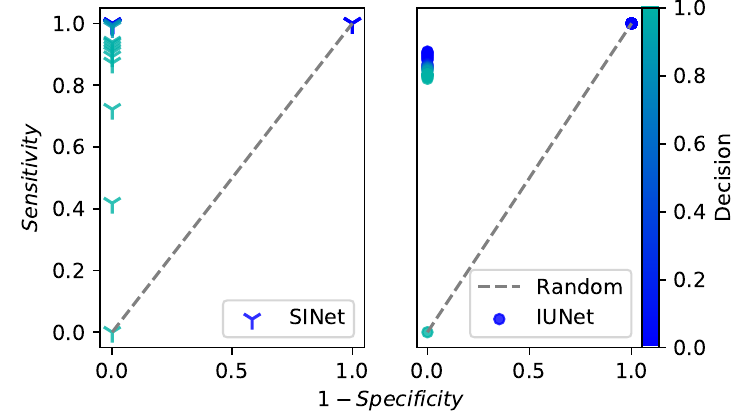}
        \label{fig:roc_test}
        }
    \caption{Receiver operator characteristic plots for volumetric (SINet) and voxel-wise (IUNet) classification.  The decision boundary represented by various points is given by the color scale.  Both models far exceed the performance of a random classifier, given by the hashed line. The AUC is 0.9977, 0.9961 in training for SINet and IUNet respectively, and 0.9996, 0.9544 in testing.}
    \label{fig:roc_curves}
\end{figure*}

The quality of the \sss module is further quantified in receiver operator characteristic (ROC) curves in Figure~\ref{fig:roc_curves}.  Given the similar behavior for $S1$ models, we only present these results using SINet as $S1$.  The ROC curve plots the true positive rate (TPR, Sensitivity, $R$) and false positive rate (FPR, 1-Specificity, FPR = $P_f/(P_f + N_t$ for true negatives $N_t$) while varying the classification decision boundary, given by the color scale.  A completely random classifier would follow the hashed line, where SINet and IUNet drastically outperform that behaviour.   These plots reinforce that SINet and IUNet are capable and accurate classifiers on both training and testing data, which is further reinforced by the area under the curve (AUC, AUROC) $AUC = 0.9977, 0.9961$ on the training dataset for $S1$ and $S2$ respectively.  The AUC is similarly high on the test dataset with $AUC = 0.9996, 0.9544$ respectively.

\begin{table}
    \centering
     \caption{Found volumes using the \sss module as designed for simulations.}
    \begin{tabular}{|c|c|c|c|}\hline
         Module     & Dataset & $z$      & $N_f$  \\\hline\hline
         SINet+S2  & PHX256-2   &  21.18   &   17       \\
         SINet+S2  & PHX-HYD    &  21.18   &   21       \\
         SINet+S2  & P3N-128    &  19.0    &   1       \\
         
         \hline\hline
    \end{tabular}
    \tablecomments{We annotate the redshift ($z$) of the output, as well as the number of volumes ($N_f$) found at that redshift. Note that P3N-128 only has SFV found at a later $z$: this is due to resolution effects of the actual simulation, highlighting a limitation of applying \sss to low-resolution data.}
    \label{tab:my_label}
\end{table}

\begin{figure}
    \subfloat[PHX256-2]{\includegraphics[width=0.45\textwidth]{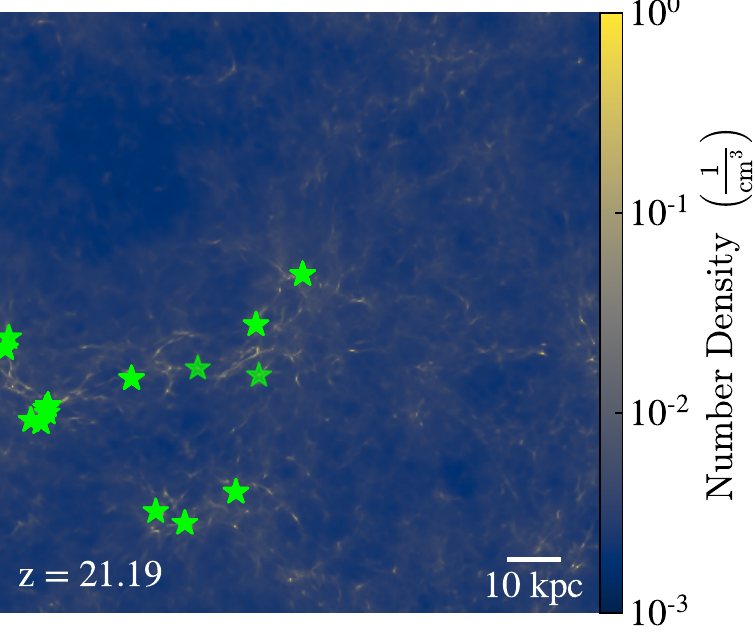}
        \label{fig:sims-256-2}}

    \subfloat[P3N-128]{\includegraphics[width=0.45\textwidth]{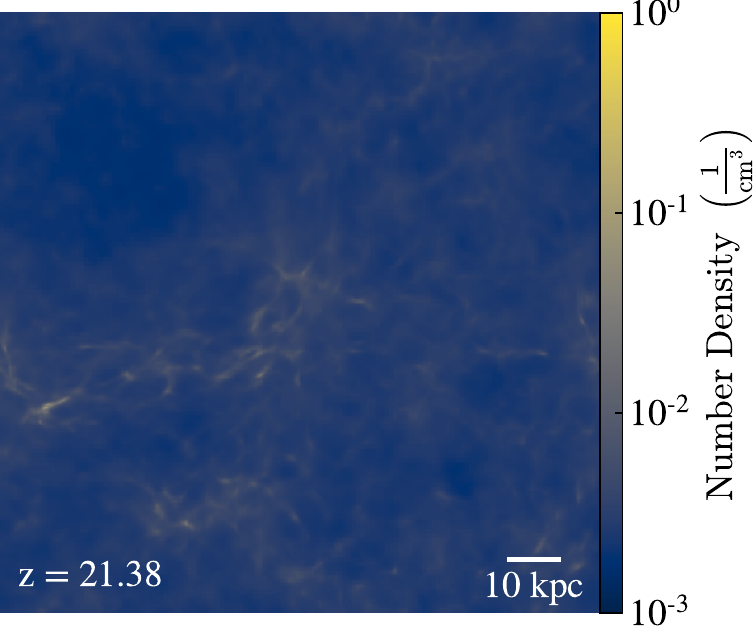}
        \label{fig:sims-128}}
    \caption{Comparing the density field of simulations with the same ICs. \ref{fig:sims-256-2} shows the PHX256-2 simulation at $z=21.19$ with \piii star forming regions annotated.  P3N-128 is shown at a similar redshift in \ref{fig:sims-128}--the density peaks in P3N-128 are both less extreme and more diffuse due to lower mass and force resolution and \sss predicts no star forming regions.}
    \label{fig:sims}
\end{figure}

\begin{figure}
    \centering
    \includegraphics[width=0.47\textwidth]{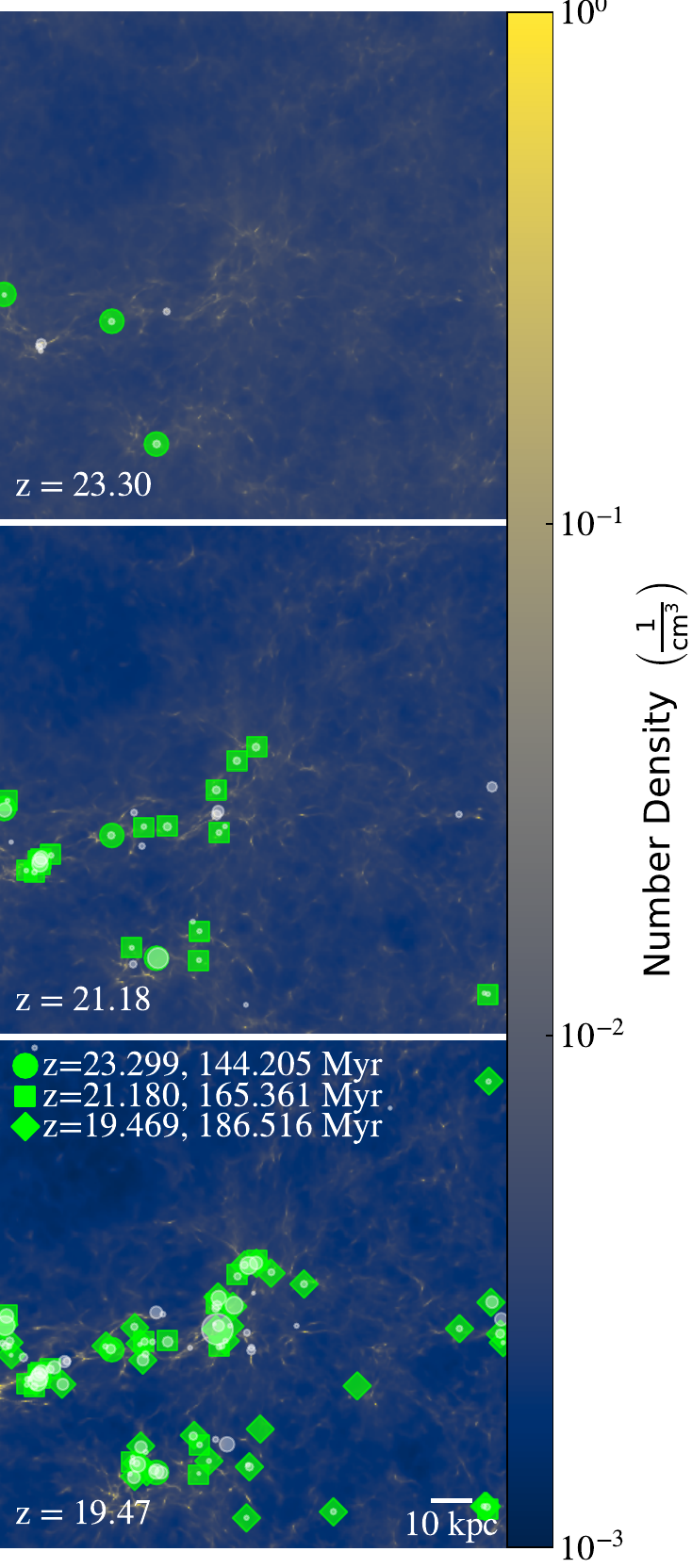}
    \caption{SFRs identified in PHX256-HYD at progressing redshifts. The middle panel is at the same redshift as PHX256-2 presented in Figure~\ref{fig:sims-256-2}.  There are more SFR predicted, e.g., at $z=21.18$, than in PHX256-2, however we have verified that the missing SFRs are present in PHX256-2 by $z=20.86$.}
    \label{fig:phx256-hyd}
\end{figure}

\subsection{Generalizability}
\label{sec:generalizability}
In all training, testing and validation data, the data was prepared identically, from simulations that have the required resolution and physical models for \piii star formation.  As a generalization test, \sss was applied to simulations that a) do not have the same physics models, and b) do not have the same resolution.  PHX256-HYD has the same cosmological parameters as the numbered PHX256 series, however features fewer AMR refinement levels (see Table~\ref{tab:test_sims}) without \piii star formation enabled.  This simulation will test the models dependence on finest-grid resolution, and since it shares ICs with PHX256-2, we would expect \sss to identify SFR in the same areas at early times, before \piii feedback has a chance to pollute the environment.  P3N-128 has half the root-grid resolution of the PHX-series simulations ($20^3$ kpccm$^3$/root-grid cell), and fewer AMR refinement levels so that $dx_{min} = 156$ pccm.  P3N-128 has no \piii star formation, and has mass resolution $\sim 1/8$ that of the PHX series.  Ten DM particles in P3N-128 have more mass than a halo expected to form \piii stars: this simulation will test how \sss performs in simulations that cannot resolve these first star forming halos.
\begin{figure*}
    \centering
    \subfloat[PHX256-HYD: $z=21.18$]{\includegraphics[width=0.32\textwidth]{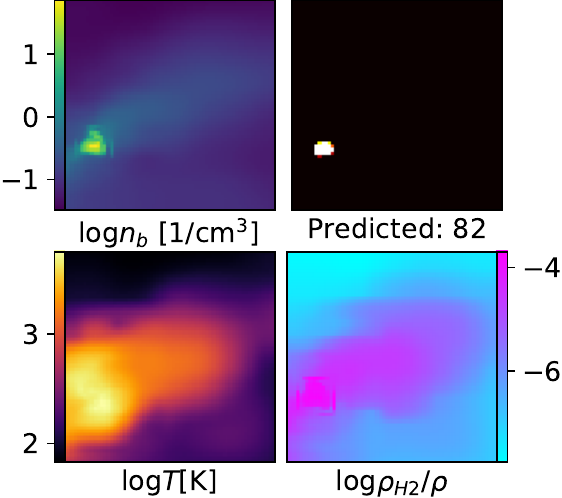}

        \label{fig:256H-1}}
    \subfloat[PHX256-HYD: $z=21.18$]{
        
        \includegraphics[width=0.32\textwidth]{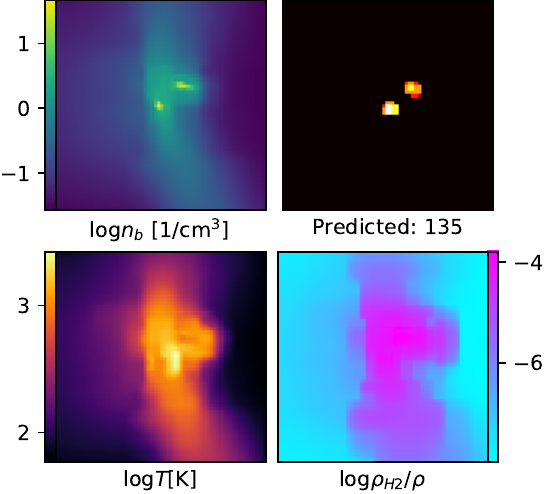}
        \label{fig:256H-2}}
    \subfloat[P3N-128: $z=19.00$]{
        
        \includegraphics[width=0.32\textwidth]{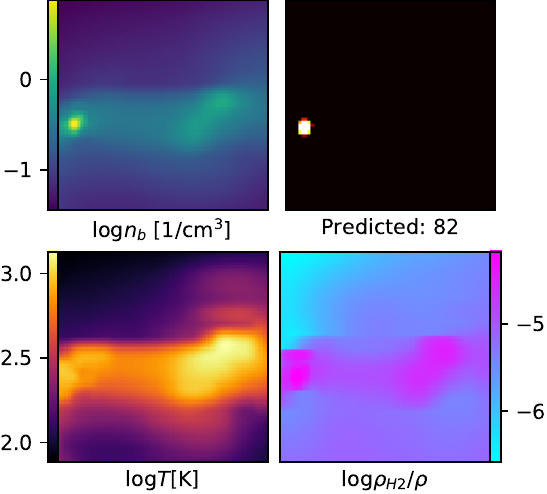}
        \label{fig:p3n-pred}}
    \caption{Test result examples from generalizability test.  Fields presented are the projection of $\log_{10}$ of baryon number density ($n_b$), temperature ($T$), and H$_2$ neutral fraction ($\rho_{H2}/\rho$); the ``Predicted'' panel shows voxels identified as SFV by $S2$ with the number of positive voxels annotated.  (a) shows a nearly ideal result with tightly clustered cells predicting a star forming region at a peak of density and H$_2$ fraction.  (b) shows the interesting possiblity of finding more than one star forming region per volume.  (c) shows identification of the first identified star forming volume (8 kpccm from that in \ref{fig:256H-1}) in an under-resolved simulation.  Since the dynamics of structure formation are less resolved, the star forming region is not found until a much later redshift.}
    \label{fig:SN128ex}
\end{figure*}

\begin{figure}
    \centering
    \includegraphics[width=0.49\textwidth]{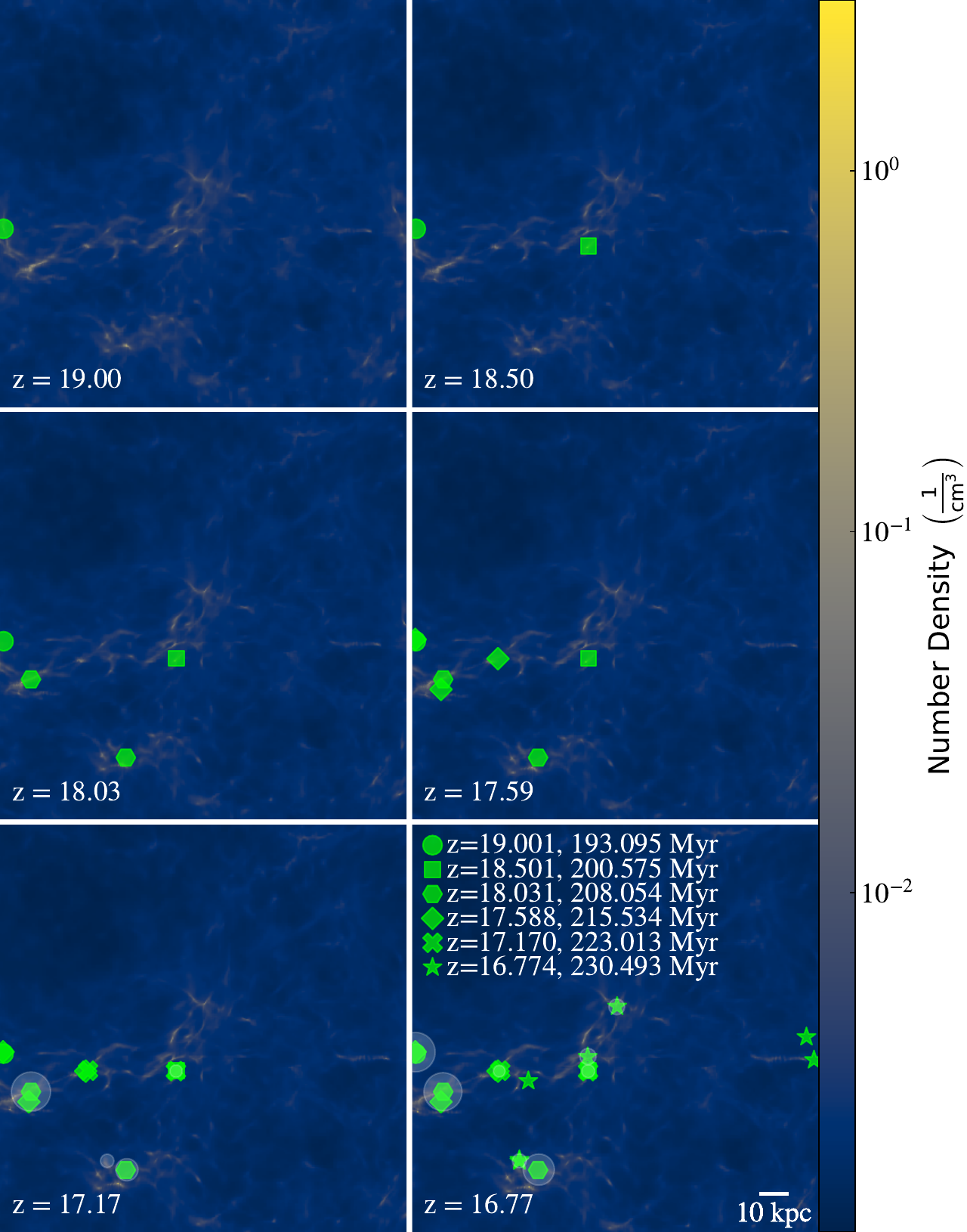}
    \caption{Redshift evolution of identified SFR in P3N-128.  Halos found by the HOP algorithm are identified by light circles (scaled by virial radius).  Most SFR are identified before their host halo is identified--by up to $\sim$37 Myr for the first SFR predicted at $z=19.0$ whose host halo is identified at $z = 16.77$.}
    \label{fig:P3N-evolution}
\end{figure}

For this test, we select a data output, and iterate through all AMR grids in the simulation hierarchy.  Any grids less than level 3 are automatically skipped, as they cannot qualify for star formation. Grids at a deep enough AMR level are tiled in 10$^3$ kpccm$^3$ volumes and each volume with $\langle \rho_B\rangle/\bar\rho > 2$ is passed through the \sss module.  If $S1$ returns nSFR, the rest of the computation is skipped.  If $S1$ returns SFR, then the volume is passed to $S2$, which will return a SFV classification for each of the $64^3$ voxels in the volume. 

In the simulations without SFF, those regions that do collapse to be identified as SFR will continue to collapse: there is no star formation to provide a sink for the gas, nor feedback to disperse or photoionize the it.  If we identify all star forming regions at, e.g., $z = 21.18$, then this will represent all the star forming regions that have formed since the simulation start.  Since the actual star formation rate (SF rate) will depend heavily on the stellar initial mass function (IMF), here we aim to match the number of star forming regions, with the assumption that a stellar IMF can then be applied to match the SF rate of higher-resolution simulations (i.e.,  PHX256).  

By $z = 21.18$, PHX256-2 has formed 17 clusters\footnote{Here, a cluster is defined as any group of stars all within 10 kpccm of the first identified star.} of \piii stars, with 226 individual star particles (see Figure \ref{fig:sims-256-2}): this early star formation is strongly clustered, averaging $>10$  star particles per 10 kpccm cluster. With a baseline for the number of star forming regions in hand from PHX256-2, we apply the module to PHX256-HYD. The results of applying \sss are presented in Figure \ref{fig:phx256-hyd}. At $z=21.18$, \sss identified 21 regions. If compared to Figure \ref{fig:sims-256-2}, we immediately identify star regions in PHX256-HYD that are not yet present in PHX256-2.  These regions however, have been verified to begin forming stars in PHX256-2 by $z = 20.86$.  \sss is identifying SFR in PHX256-HYD at locations reflected in PHX256-2, and all of the regions identified by \sss are mirrored by or precede star formation in PHX256-2.  
 
Since there is no feedback implemented, a full analysis of errors incurred by using \sss and its feedback method is deferred to future work.  Of primary interest will be the errors in star forming locations and the metal distribution resulting from \piii stars, however we will also investigate, e.g., the impact of early or late star formation predictions such as those seen in the PHX256-HYD.

Example projections of the predictions from our generalization tests are shown in Figure~\ref{fig:SN128ex}. Shown are fields that correspond more easily to physical intuition than the fields which generated the predictions; we show number density $n_b$ in place of baryon density, temperature, $T$, in place of energy, and H$_2$ neutral fraction ($\rho_{H2}/\rho_ b$) in place of neutral H$_2$ density.  Figure~\ref{fig:256H-1} shows an ideal prediction; the predicted region is compact and well defined, and appears to agree with peaks in both $n_b$ and $\rho_{H2}/\rho_ b$, which are both to be expected from the star formation algorithms in \enzo. Figure~\ref{fig:256H-2} shows the prediction of two star forming regions in one volume.  While examples like this are possible, they become more common as the collapse dynamics proceed without stellar feedback and enriched star formation to begin disqualifying regions for \piii star formation.  The dual predictions still coincide with sensible locations in the projected fields.  Figure~\ref{fig:p3n-pred} shows the first prediction of SFR in P3N-128, in a neighboring region near \ref{fig:256H-1} at $z=19.0$.  The relative lateness of this prediction is an artifact of structure formation dynamics in under-resolved simulations; as seen in Figure~\ref{fig:sims-128}, at $z\simeq21$, the density field in P3N-128 is amorphous, showing no sharp, distinct features as seen in the higher resolution simulations.

In applying the \sss module to P3N-128, the module identifies no star forming regions at $z=21.38$.  To analyze how star formation as identified by \sss may proceed, we iterate through all available outputs of P3N-128 to find both the first SFR, and subsequent SFR.  The results of this test are presented in Figure~\ref{fig:P3N-evolution}.  The first SFR is identified at $z=19$, with more found at each subsequent output.  Figure~\ref{fig:P3N-evolution} does not include every SFR--those that are very close together are only plotted once to aid readability: at $z=16.77$, there are 17 distinct SFR identified in P3N-128.  This raw number of SFR and their locations in the projection agrees well with those in PHX256-2 at $z=21.19$, as seen in Figure~\ref{fig:sims-256-2}.

\section{Discussion}
\label{sec:disc}
\subsection{Further discussions of generalizability}

\indent The \sss module has one major goal: to identify \piii star forming regions in simulations without the necessary resolution for star formation.  The lack of resolution could present itself in a spatial sense, as in PHX256-HYD, or in a mass sense, as in P3N-128. We find that the module cannot completely compensate for the lack of mass resolution in P3N-128--the density peaks that would lead to star formation are evolving at a different rate and in a different manner, with peaks being more broad and potential star forming regions being less well-defined than a similar redshift in PHX256-2.  Figure~\ref{fig:sims} shows a comparison between PHX256-2 and P3N-128.  Even by eye, the dynamics of collapse in \ref{fig:sims-128} are very different than \ref{fig:sims-256-2}, with density peaks being both less extreme and more diffuse.   
\begin{figure}
    \centering
    \includegraphics[width=0.49\textwidth]{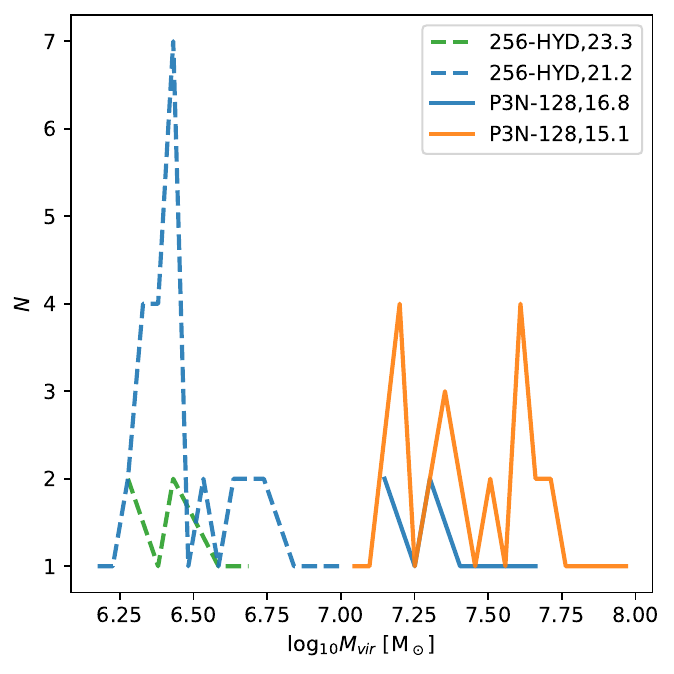}
    \caption{Halo number ($N$) as a function of log halo virial mass comparison between PHX256-2 and P3N-128 at redshift annotated in the legend.   Note that halo formation in P3N-128 begins at $z\lesssim 17.17$, and low-mass, \piii forming halos ($M_{vir} \gtrsim 10^{6.5}$~\msun) are never formed in P3N-128 due to its low mass resolution.}
    \label{fig:hmf}
\end{figure}

 We can further illustrate the effect of structure formation from a halo mass function (HMF) perspective.  Given that PHX256-2 and P3N-128 share identical initial conditions and box size, then without resolution effects, we should expect that they have the same HMF at a given redshift.  To get the HMF (represented here by halo count as a function of halo mass), we use the HOP algorithm~\citep{eisenstein1998} with overdensity threshold 100 to identify DM halos within both simulations. Comparisons of halo number are shown in Figure \ref{fig:hmf}.  At $z\simeq 21$, P3N-128 has no identifiable halos;  the first ones start to appear with virial mass $M_{vir} \sim 10^{7.2-7.6}$~\msun~at $z=17.17$.  P3N-128 never creates low mass halos with $M_{vir}\sim 10^{5-6.5}$ that would be expected to form \piii stars, as the particle mass is simply too high and those halos are unresolved.  Despite this lack of resolution, as presented in Figure~\ref{fig:P3N-evolution}, \sss predicts stars that will be in halos that form 10-30 Myr after the SFR identification. 
 
 The mass (particle mass or gas mass in a grid cell) is a determining factor in the effectiveness of stellar feedback ~\citep{Hopkins2019FB}.  To determine the potential difference of feedback applied at $z=19$ or $z=16.77$, we analyzed the mass distribution of the SFR ($M_{SFR}$) as compared to the later halo ($M_{halo}$).  We found that $M_{SFR} < 0.5M_{halo}$, when using the virial radius ($R_{vir}$) of the halo at $z=16.77$ to define a volume at both redshifts.  Importantly, the gas is less compact in the central regions e.g., at $R=R_{vir}/7$, $M_{SFR} < 0.2M_{halo}$.  When we extend this analysis to all SFR identified by $z = 17.59$, comparing each SFR to the later halo it is nearest to, we find that the SFR has overdensity  $[\langle\rho\rangle/\bar\rho]_{SFR} \lesssim 0.5 [\langle\rho\rangle/\bar\rho]_{halo}$.  If halos are too massive when they start forming \piii stars, the feedback will be confined and unable to pollute the local environment~\citep{Whalen2008b}, so applying feedback at the less dense states identified by \sss may reduce the effect of the massive halos found in P3N-128.

\subsection{Error Analysis}
One of the primary concerns when applying a deep learning model to a production application is to understand how the model fails.  If the failure modes are predictable, then they can be accounted for in the deployment of the model, but if they are seemingly random, the model may not even be usable.  First, we are concerned with failures to classify in a volumetric sense, i.e., identifying ``nearby" star formation is acceptable, but misclassifying an entire region is a much more egregious failure. To analyse these critical failures of the \sss module, we passed every training/test/validation sample through the model, and plotted a) the SFV labels if $S1$ was incorrect, or b) the SFV predictions and labels if $S2$ incorrectly classified the volume.  65\% of the false positives in $S1$ were cases where the star voxels were within 3 voxel-widths of the border of the volume.  All false positive cases of $S2$ were on the edge of the volume, and 75\% of the false negatives had star forming labels on the border.  The star forming border is a major avenue of failure for the module, but also an easy one to remedy.  In a production pipeline, two obvious possibilities could alleviate these failures: a) ignore the SFR if the flagged voxels are on the edge, as the routine will be called again in $< 1$ Myr, at which point the SFV will have likely moved to a more central location, or b) recenter the volume on the suspicious SFV and re-run the module to receive a more reliable result.  The results in this work used the second option to reduce false positives and negatives in analysis of the PHX256-HYD and P3N-128 simulations. 

\begin{figure}
    \centering
    \includegraphics[width=0.45\textwidth]{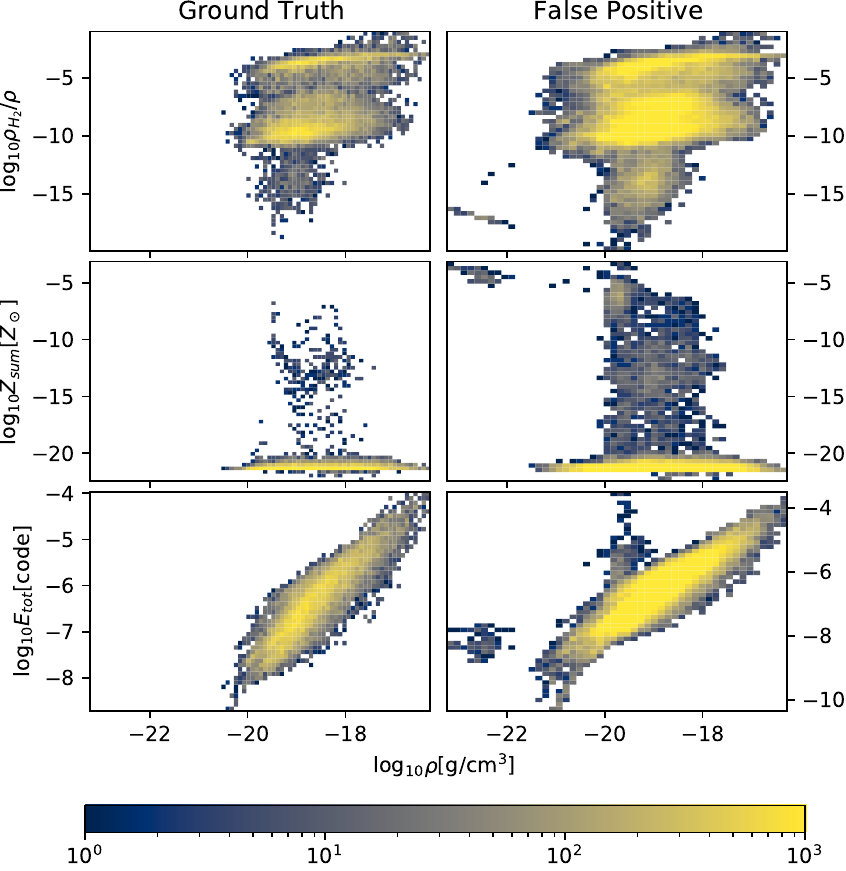}
    \caption{Voxel properties of ground truth voxels and false positive predictions. We present 2D histograms of $\log_{10}$ density ($\rho$) with other hydrodynamic quantities: $\log_{10}$H$_2$ fraction, $\log_{10}$ total metallicity ($Z_{sum}$), and $\log_{10}$ total energy ($E_{tot}$) from top to bottom. While false positives qualitatively fall into a similar distribution as the ground truth, there are obvious outliers.}
    \label{fig:voxelstats}
\end{figure}

We can additionally examine the predictions of \sss by comparing voxel-based quantities for ground truth and false positive voxels.  We collected the voxel field quantities for positive predictions within the testing data, to generate Figure~\ref{fig:voxelstats}, which shows 2-D histograms of $\rho-$H$_2$ fraction (top), $\rho-Z_{sum}$ (middle), and $\rho-E_{tot}$ (bottom), where the total energy given is quoted in code units and metallicity is given in $Z_\odot$. Qualitatively, the distributions look similar, however there are visible artifacts in the false positive (right) panels.  Also visible in the $\rho_{H_2}$ plot is a bimodal distribution, which is an artifact of labelling a $3^3$ cloud as star forming, as opposed to including only the single voxel that hosts the star particle.  Distributions such as those in Figure~\ref{fig:voxelstats} can be used in the future to check that the predictions from \sss are reasonable, or to reduce false positives simply by screening out voxels that fall far outside the distribution from training data.  We can additionally use the diffuse points in the ground truth distribution to identify where the prototype training/test dataset is insufficient.  For example, examination of the $\rho-Z_{sum}$ ground truth histogram (middle-left) shows very few voxels with finite but low metallicity ($Z > 10^{-10}$), where the false positive panel shows a significant number ($\gtrsim10^3$) with $10^{-10} < Z_{sum} < Z_{crit}$.  Focusing future dataset generation on these poorly sampled spaces will increase the robustness of the final model.  Guided by Figure~\ref{fig:voxelstats}, we tested a simple screen that requires any SFV to have $n_b > 1.0$.  Such a simple filter, on average, reduced false positive voxels by 11 and decreased the $\langle IoU\rangle$ by 0.02, while having no effect on false negative or true positive predictions.  Future work will increase the robustness of \sss by increasing training samples from poorly-sampled regions and quantifying the failures so that they can be compensated for in a real-time simulation.

\section{Conclusions}
\label{sec:conclusions}

We have designed a classification-segmentation model using deep ANN that is capable of predicting \piii star formation sites in cosmological simulations. This is the only method known to the authors that can accomplish such a feat.  Our findings can be summarized as follows:
\begin{enumerate}
    \item We have found that well-known image recognition architectures (adapted to 3D) are effective at this task, as are common pixel segmentation architectures.  The choice of architecture had little impact on classification capability given the small classifiers developed here.
    \item The \sss module predicts 9.98$^3$ kpccm$^3$ star forming volumes with $>99.8\%$ accuracy, reinforced by volumetric classification completeness, TSS, HSS, precision, and recall. 
    \item The module has been applied to hydrodynamic simulations that have no star formation routines enabled and predicted star formation in the same regions at similar redshift as high resolution full-physics simulations. 
    \item In many cases, \sss predicts star formation regions well before the formation of the host halo, particularly in under-resolved simulations such as P3N-128. More testing will be needed to quantify the success of \sss in those cases.
    \item By many classical measures of classification algorithms, i.e., ROC, AUC, precision, recall, $F_1$, and completeness, \sss is a capable classifier in a volumetric sense.  Correctly classifying volumes is much more important (in our application) than predicting the exact location of star formation.
    \item Voxel-wise predictions IUNet suffers from a high number of false positive voxel predictions, as indicated by low precision.  It over predicts the size of localized regions in SFR, which is reinforced by the high $\langle IoU\rangle$ and simple statistical counting of SFV. Predicted star forming regions appear well defined (e.g., Figure~\ref{fig:SN128ex}), but cover a larger region than exists in the ground truth. The predicted region still generally covers the ground truth SFV, as indicated by the high value of the voxel-wise recall in Table~\ref{tab:s3results}.
    \item Cosmology and star formation simulations are inherently multi-scale problems, and we hypothesize that the multi-scale sensitivity of IBs is contributing to the strong performance of IUNet, as well as SINet being quantitatively as capable as SDNet and SRNet, despite having $\sim 20 \times$ fewer trainable parameters.
    \item Utilizing voxel statistics from our early datasets is one way to increase the quality of the $S2$ classification without further training or expensive calculation, or to guide further dataset generation as the PHX-series simulations progress to lower redshift.
    \item Despite these successes, fully quantifying the success of the model by comparing model-assisted simulations with ground-truth conventional sub-grid simulations will require an active stellar feedback method so that regions can move beyond primordial stars and into enriched stellar population production.
    
\end{enumerate}

\sss as a whole may be improved most by increasing the reliability and robustness of $S2$. This is the primary goal of on-going development; it may be improved through testing different loss functions, a combination of loss functions~\citep{Taghanaki2018, Hajiabadi2020}, or improving the representation of poorly sampled cases, as seen in Figure~\ref{fig:voxelstats}.  Future work will also focus on analyzing the failures of \sss so that they can be more accurately estimated and guarded against.

This method will ultimately be used in a novel sub-grid feedback method, where SFR identified using this algorithm will then be evolved to a post-star state using another series of deep neural networks. The goal throughout this work was to design the proof of concept model that could be used in a production capacity (further training and development not withstanding): to that end, here we estimate the run-time implications of using this module to perform \piii SFF.  Evolving PHX256-2 from z=21.19 to z=17.53 consumed 7K cpu-hours. Most of this time is spent evolving the hydrodynamic and radiation fields from hot supernova remnants or \piii main sequence stars in the deep AMR levels of the volume.  PHX256-HYD evolves the same region in 14 cpu-hours by excluding \piii star formation and restricting the maximum AMR level.  Applying the \sss module to locate star forming regions once (i.e., one panel in Figure \ref{fig:phx256-hyd}) consumes $\sim$4 CPU-hours, processing 5.2 volumes/sec on average. To localize \piii star forming regions every Myr would require performing this evaluation $\sim 30$ times, consuming a total of 120 CPU-hours.  Most of the \sss processing time is spent iterating through grids and disqualifying regions before they even enter $S1$: only $\lesssim5$\% of potential regions are classified by $S1$, with 3\% continuing to $S2$, where only 2\% finally qualify as SFR (0.003\% of potential volumes), so we estimate that adding a feedback routine after $S2$ would not significantly increase the processing time of the algorithm as a whole.  This method, when finalized with a feedback algorithm, may produce $\gtrsim50\times$ speedup as compared to explicitly star forming simulations, with more speedup likely from optimizing the final method.

This rate makes this method feasible to use inline with an \enzo  simulation using inline python with {\tt YT}, particularly since the tests run here do not include optimization of any kind.  We could expect further speedup by incorporating the method into \enzo's source C/C++ code and/or optimizing \sss using MPI-parallelism on CPU architecture (e.g., \citealt{mathuriya2018}).  Future work will focus on evolving \piii stellar remnants in regions identified using this method, and incorporating the framework into an \enzo  simulation, along with further development and training of the \sss module.

\bigskip
\noindent
This research was supported by National Science Foundation grant CDS\&E grant AST-1615848 to M.L.N. The simulations
were performed using \enzo ~on the Frontera supercomputer operated
by the Texas Advanced Computing Center (TACC)
with LRAC allocation AST-20007. Data analysis and model development, training, and testing were performed on the
Comet supercomputer operated for XSEDE by the San Diego
Supercomputer Center under XRAC allocation TG-AST200019. 

\bibliographystyle{aasjournal}
\bibliography{bib}
\end{document}